# Simulation and experimental observation of tunable photonic nanojet and photonic hook upon asymmetric illumination of a mesoscale cylinder with mask

Igor V. Minin,[1,2] Oleg V. Minin,[1,2] Cheng-Yang Liu,[3,*] Hao-De Wei,[3] Alina Karabchevsky[4]

[1]*Tomsk Polytechnic University, Tomsk, 634050, Russia*
[2]*Tomsk State University, Tomsk, 634050, Russia*
[3]*Department of Biomedical Engineering, National Yang-Ming University, Taipei City, Taiwan*
[4]*School of Electrical and Computer Engineering, Ben-Gurion University of the Negev, Beer-Sheva, 8410501, Israel*

**In this letter, we report on numerical study, fabrication and experimental observations of tunable photonic nanojet and photonic hook. Here, the curved photonic nanojets (photonic hooks) are generated by single mesoscale microcylinder that we fabricated from polydimethylsiloxane (PDMS), upon its boundary illumination and adjustable area at an incident wavelength of λ = 405 nm. Experimental observations were conducted for PDMS microcylinder of diameter *d* = 5 μm deposited on a silicon substrate with the aluminum mask. Measurements were performed with a scanning optical microscope. Our experimental results are in good agreement with numerical predictions performed with the finite-difference time-domain method. The observed the full widths at half-maximum of photonic hooks are 0.48λ, 0.56λ, and 0.76λ for tilt angles of *θ* = 0°, 5.7°, and 20.1° respectively, at the mask heights of *h* = 0, 0.25*d* and 0.5*d*, respectively, displaying the specificities of the field localization. These photonic hooks have great potential in complex manipulation such as super-resolution imaging, surface fabrication, and optomechanical manipulation in curved trajectories smaller than wavelength. 2020**

From ancient times, it was well-known that light propagates in straight lines named rays [1]. However, in 2007, the self-accelerating Airy beams which do not propagate as rays but rather curved beams were proposed and experimentally discovered in optics [2,3]. The Airy-like accelerating curved beams exhibit many very important features [4]. In 2015, an original type of subwavelength curved beam named photonic hook (PH) was proposed based on the physical principles of photonic nanojet (PNJ) formation [5-7]. The optical properties of PNJ are a function of particle geometry, refractive index contrast, and dimensions [8-10]. Moreover, a dielectric mesoscale particle must have Mie size parameter $q$ corresponding to $q \sim (2...20)\pi$, where $q = 2\pi r/\lambda$, $r$ is the radius of the particle and $\lambda$ is the incident wavelength.

As a simple example of PH, a Janus particle in the form of a cuboid with broken symmetry was studied and experimentally verified [11-13]. It is important to note that PH has a radius of curvature, a lateral size of subwavelength, and no curved side lobes [14]. The influence of specific illumination conditions (coaxial illumination with the adjustable area and boundary illumination) for spherical $BaTiO_3$ particles with a diameter of approximately 127λ ($q \sim 42\pi$) [15] and $SiO_2$ microspheres with a diameter of 433λ ($q \sim 144\pi$) [16] (which are however out of mesoscale conditions) was considered in the optical waveband. This phenomenon is similar to a cylindrical lens with a decentered aperture under normal illumination because of the large spherical aberration [17,18]. The formation of the PH through a glass cuboid embedded in a structured dielectric cylinder was studied numerically [19]. Furthermore, two PHs were observed by the specially designed five-layer dielectric cylinder [20]. Most of these known solutions usually require specially designed particles or complex processing methods, which limits the further development of this type of curved beams. The direct experimental confirmation of the curved focus is still not performed under the conditions of small mesoscale particles. In this paper, we experimentally report the direct imaging of curved PNJs created by a mesoscale particle of cylindrical shape ($q \sim 12\pi$) while partially blocking an incident light by an amplitude metal mask. For this, we study the key parameters of the PH such as lateral full width at half maximum (FWHM), tilt angle, and maximal intensity depending on the variation of the mask height. We fabricated the designed microcylinder and characterized it to verify its focusing properties.

The schematic of the studied system with the definition of curvature for a PNJ is shown in Fig. 1(a). Figure 1(b) shows the artistic representation of the system generating curved localized beams. We studied numerically the production capability of curved

PNJ using a dielectric microcylinder illuminated by a laser beam under different height $h$ of the metallic mask. This laser beam is a quasi-plane wavefront since the Gaussian beam with a breadth of 1 mm is much bigger than 5 μm diameter of dielectric microcylinder. We illuminate the cylinder with the incident wavelength λ = 405 nm. The metallic mask is placed next to the cylinder at a distance of less than 1 μm. The curvature of the PH is defined by the tilt angle $θ$ [11-13]. To simulate the near-field structure of localized electromagnetic waves, we use the finite-difference time-domain (FDTD) method with perfectly matched layer (PML) boundary conditions [21]. We have built several numerical models of 5 μm microcylinder coupled with a metallic mask at different $h$ values for quantifying the influence of the illuminating beamwidth.

The wave vector $K_∥$ is relative to the axis of symmetry of the cylinder which creates the PNJ curvature profile. On the other hand, the components of the wave vector $K_⊥$ determine the length of the PNJ along the propagation direction. Therefore, the local interference of the optical fields inside the cylinder can generate the PH. This will depend on the width of the illuminating beam. A similar effect can be observed while using the internal asymmetry of the particle material (Janus particles) with a full illuminating beam [24].

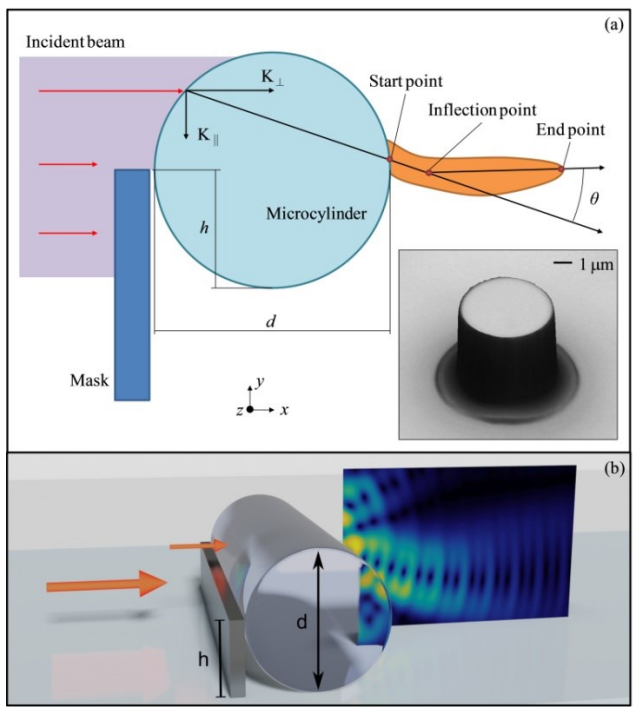

Fig. 1. (a) Schematics description of curvature for a photonic hook. The insert shows the microphotograph of the 5 μm dielectric microcylinder. (b) Schematic render of the studied system.

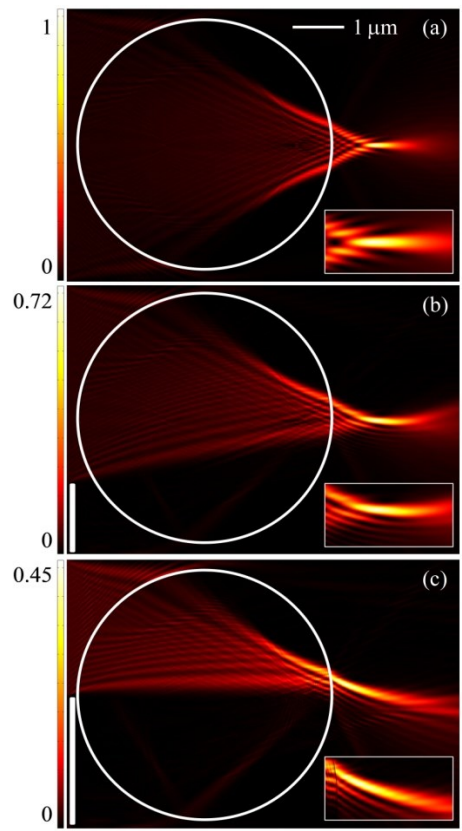

Fig. 2. Numerical results of normalized intensity maps for the dielectric microcylinder with metallic mask at (a) $h = 0$, (b) $h = 0.25d$, and (c) $h = 0.5d$. The incident wavelength for these calculations and experiments is 405 nm. Zoomed in PH is shown in insets.

To reveal the curvature changes of the PHs, the field intensity distributions at several key $h$ values by numerical simulations are illustrated in Fig. 2. The case of $h = 0$ corresponds to the classical PNJ for dielectric microcylinder. As shown in Fig. 2, the PH's shape and curvature radius can be adjusted by varying mask height. In particular, when comparing Figs. 2(a)-2(c) we noticed, that the curvature of the PH changes when the mask height varies from $h = 0$ to $0.5d$. The physics of the curved PNJ formation can be explained as follows. The refractive index of cylinder and width of illumination determines the angle of refraction on an interface via the generalized Snell's law [22]. Part of the illuminating beam, determined by the mask height, is refracted first on the front surface of the cylinder. Then, the light beam inside the cylinder is refracted a second time when it exits from the shadow surface of the cylinder (see Fig. 1(a)). If the width of the illuminating beam is less than the cylinder diameter, the components of the wave vector $K_∥$ do not cancel each other through the local destructive interference [23].

To experimentally demonstrate the PH phenomenon based on spatially limited illumination, we built a setup shown in Fig. 3. It consists of a linearly polarized single-mode diode-pumped solid-state laser. The variable density filter is used to modify the intensity of the incident laser beam. The metallic mask and substrate (silicon wafer with refractive index $n_w$ = 5.567+0.386i) [25] are mounted on two independent stages. The linear motorized stages (SigmaKoki SGSP20-85) have 100 nm lateral resolution in both $x$ and $y$ directions for positioning the metallic mask and substrate. The objective with high numerical aperture is mounted on a piezoelectric actuator (SigmaKoki SFS-OBL-1) with 10 nm resolution in the $z$-direction. The piezoelectric actuator in the $z$-direction was used to define the focal plane of interest with the precision of tens nanometers and successfully imaged the PH and PNJ via a CMOS camera (Whited UC-1800). To avoid multiple internal reflections between the microcylinder and other elements, the laser beam illuminating the microcylinder is absorbed by a beam dump. All imaging system is located in a light-controlled darkroom for preventing any effect of the background noise.

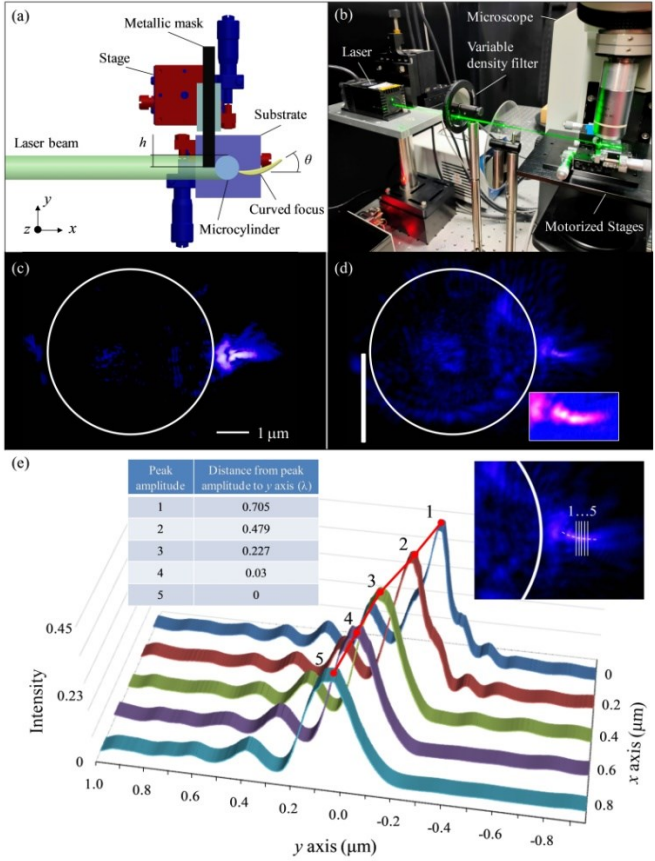

Fig. 3. (a) Schematic diagram of the curved focus generated by a dielectric microcylinder with a metallic mask. (b) Experimental configuration of the scanning microscope system. Experimental raw images of intensity maps for the dielectric microcylinder with metallic mask at (c) $h = 0$, and (d) $h = 0.5d$. The incident wavelength is 405 nm. The insert indicates the enlargement of PH. (e) Experimental cross-sections of the observed curved focus at $h = 0.5d$. The insert indicates the positions of serial cross-sections. The table-inset shows the offset distance from each peak amplitude to $y$-axis.

We fabricated the dielectric microcylinder with refractive index $n = 1.41$ and a height of 6 μm out of polydimethylsiloxane (PDMS) using conventional photolithography and replica molding process [26]. The PDMS microcylinder was then placed on the silicon wafer. The material of the metallic mask is aluminum and the mask width is about 100 μm. A laser scanning digital microscope (LSDM) is used to measure the surface profile of the PDMS microcylinder [27]. The LSDM image of a single PDMS microcylinder is presented in the inset of Fig. 1(a). The top and the bottom diameters of the microcylinder were measured as 4.92 μm and 5.31 μm respectively. The value of surface roughness for this microcylinder was evaluated as Ra = 16 nm which indicates a reasonable uniformity in the diameter of PDMS microcylinder. We concluded that the PDMS microcylinder with a mask is suitable for generating PH. In the experiment, the PDMS microcylinder is illuminated by a coherent laser beam along the $x$-direction which is shown in Fig. 3(a). We assemble the scanning optical microscope to capture the experimental raw images [28]. The scanning in the $z$-direction is performed for obtaining clear raw images of the PHs [29]. Figures 3(c) and 3(d) show the experimental results of PH visualization. The comparison of the results from Figs. 2(c) and 3(d) we notice that the intensity distributions for the numerical and experimental results are in good agreement. In both simulations and experiments, the variation of PH curvature depends on the mask height. Importantly, by tuning the mask height, the curvature of PNJ is tuned in the $y$-direction.

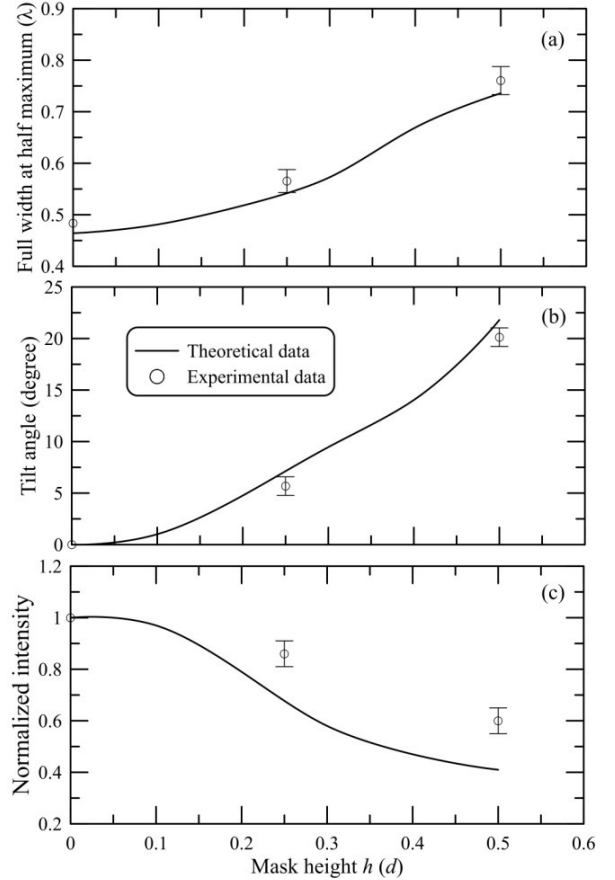

Fig. 4. (a) Full width at half maximum, (b) tilt angle, and (c) normalized intensity of the curved focus as a function of the mask height.

Figure 4 summarizes the PH curvature and the peak intensity compared to these in the full illumination case ($h = 0$). To evaluate the FWHM of PH, we find the start point and the inflection point (peak amplitude) in the experimental images. The propagation direction is from the start point to the inflection point as shown in Fig. 1(a). The FWHM is the double distance perpendicular to the propagation direction between the maximal peak amplitude and the half-maximum point. The tilt angle $\theta$ is identified as the angle between the propagation direction and horizontal axis ($x$-axis). Figures 4(a) and 4(b) show the FWHM and the tilt angle of the produced PH at the point of the peak amplitude, respectively. The diffraction-limited PH (FWHM $\leqq \lambda/2$) is observed at mask height $h \leqq 0.15d$ with a small tilt angle less 2.5° while the maximal field intensity decreases by up to 10%. However, the FWHM of the PH is more than that of a classical PNJ at $h > 0.15d$ as well. The mask height has a significant influence on the curvature of the PH. When mask height increases, the tilt angle is on a rising trend. It can be seen that the tilt angle $\theta$ of the PH is about 20° at the mask height of $h = 0.5d$. The full length of the PH in the case of $h = 0.5d$ is about 3 μm in which the distance between the origin and inflection point is 0.8 μm and the distance between the inflection point and the target point is 2.2 μm. In case there is no mask, the FWHM of the PNJ is less than half of the wavelength. However, the FWHM of the PH is 0.76λ when the mask height is equal to half of the cylinder diameter. In Fig. 4(c), the peak intensity of the PH decreases as the mask height increases. At $h = 0.5d$, the peak intensity has decreased to approximately 0.5, half of its original intensity.

To demonstrate the curvature of the PH, the experimental cross-sections of the observed curved focus at $h = 0.5d$ are shown in Fig. 3(e). Serial transverse cross-sections correspond to serial focal planes along the *x*-axis moving upwards by steps of 200 nm. We acquire the cross-sections from the raw images in Fig. 3(d) and select 5 cross-sections from the inflection point for better observation, as schematically shown in the inset of Fig. 3(e). In general, the classical PNJ is formed from a dielectric microcylinder (see Fig. 2(a)). When the incident beam illuminates a full surface of the microcylinder, the optical beam will be divided into two branches inside the microcylinder and converge into a PNJ near the shadow surface of the microcylinder [6,7]. For a masked microcylinder, one of the branches inside the microcylinder is blocked by the metallic mask as shown in Fig. 2(c). Only one part of the optical beam inside the microcylinder is refracted on the rear surface of the microcylinder, and then a curved focus area is formed near the shadow surface of the microcylinder. Another key factor to remember is that the wave oscillations of a complete phase cycle inside the microcylinder are irregular due to the asymmetry of illuminating waves by mask apodization and this is the cause of an optical bending beam. The table-inset in Fig. 3(e) shows the offset distance from each peak amplitude to the *y*-axis. It can be seen that the position of the peak amplitude in the region of radiation localization shifts in the transverse direction (*y*-axis) when the distance from the shadow surface of the microcylinder increases in the horizontal direction (*x*-axis). In other words, the localized optical beam is not direct and the focusing beam is bent to form a PH near the rear surface of the microcylinder. Analysis of experimental data shows that the offset distance of the PH increases 5.5 times in the 1 μm length with increasing mask height from $h = 0.25d$ to $h = 0.5d$. Thus, increasing the mask height leads to a significant increase in the PH curvature. In terms of tilt angle, a larger $h$ brings a bigger $\theta$. Compared to a conventional PNJ, the PH should lead to an increase in spatial resolution and field of view. However, the price of increasing the PH curvature is of the maximum intensity of the focus, as shown in Fig. 4(c).

In conclusion, we directly observed experimental images of curved PNJs in the visible light region. In contrast to known methods of PH generation, the PH can be simply created using a mesoscale cylindrical dielectric particle with an amplitude mask. The PH generation considered in this letter does not require the manufacture of microparticles with a special shape or complex internal structure. The mesoscale dimensions and simplicity of the PHs are much more controllable for practical tasks and indicating a wide range of potential applications. Finally, we would like to stress that the observed method of PH generation should be inherent to acoustic and other beams including surface waves and microwaves for interacting with mesoscale symmetric obstacles and asymmetric illumination.

**Funding.** Ministry of Science and Technology, Taiwan (MOST) (109-2923-E-010-001-MY2); Yen Tjing Ling Medical Foundation (CI-109-24); Israeli Innovation Authority funding (69073); Russian Foundation for Basic Research (20-57-S52001)

**Disclosures**. The authors declare no conflicts of interest.

**Author contributions**. I.M. and O.M. conceived the idea, wrote the manuscript and supervised the project. C.-Y.L. and H.-D.W. performed the experiments. C-Y.L. performed simulation and wrote the manuscript. A.K. edited the manuscript and figure. All the authors revised the results and the manuscript.

## References

1. L. Novotny and B. Hecht, *Principles of Nano-Optics*, (Cambridge University, 2006).
2. G. Siviloglou and D. Christodoulides, "Accelerating finite energy Airy beams," Opt. Lett. **32**, 979-981 (2007).
3. G. Siviloglou, J. Broky, A. Dogariu, and D. Christodoulides, "Observation of accelerating Airy beams," Phys. Rev. Lett. **99**, 213901 (2007).
4. N. Efremidis, Z. Chen, M. Segev, and D. Christodoulides, "Airy beams and accelerating waves: an overview of recent advances," Optica **6**, 686-701 (2019).
5. I. V. Minin and O. V. Minin, *Diffractive Optics and Nanophotonics: Resolution Below the Diffraction Limit* (Springer, 2016).
6. A. Heifetz, S. Kong, A. Sahakian, A. Taflove, and V. Backman, "Photonic nanojets," J. Comput. Theor. Nanosci. **6**, 1979-1992 (2009).
7. B. Luk'yanchuk, R. Paniagua-Domínguez, I. V. Minin, O. V. Minin, and Z. Wang, "Refractive index less than two: photonic nanojets yesterday, today and tomorrow," Opt. Mater. Express **7**, 1820-1847 (2017).
8. C. Liu and F. Lin, "Geometric effect on photonic nanojet generated by dielectric microcylinders with non-cylindrical cross-sections," Opt. Commun. **380**, 287-296 (2016).
9. C. Lin, Z. Huang, and C. Liu, "Formation of high-quality photonic nanojets by decorating spider silk," Opt. Lett. **44**, 667-670 (2019).
10. C. Liu and M. Yeh, "Experimental verification of twin photonic nanojets from a dielectric microcylinder," Opt. Lett. **44**, 3262-3265 (2019).
11. L. Yue, O. V. Minin, Z. Wang, J. Monks, A. Shalin, and I. V. Minin, "Photonic hook: a new curved light beam," Opt. Lett. **43**, 771-774 (2018).
12. A. Ang, A. Karabchevsky, I. V. Minin, O. V. Minin, S. Sukhov, and A. Shalin, "Photonic Hook based optomechanical nanoparticle manipulator," Sci. Rep. **8**, 2029 (2018).
13. I. V. Minin, O. V. Minin, G. Katyba, N. Chernomyrdin, V. Kurlov, K. Zaytsev, L. Yue, Z. Wang, and D. Christodoulides, "Experimental observation of a photonic hook," Appl. Phys. Lett. **114**, 031105 (2019).
14. K. Dholakia and G. Bruce, "Optical hooks," Nat. Photonics **13**, 229-230 (2019).
15. F. Wang, L. Liu, P. Yu, L. Zhu, H. Yu, Y. Wang, and W. Li, "Three-dimensional super-resolution morphology by near-field assisted white-light interferometry," Sci. Rep. **6**, 24703 (2016).
16. E. Xing, H. Gao, J. Rong, S. Khew, H. Liu, C. Tong, and M. Hong, "Dynamically tunable multi-lobe laser generation via multifocal curved beam," Opt. Express **26**, 30944-30951 (2018).
17. Z. Cao, C. Zhai, J. Li, F. Xian, and S. Pei, "Light sheet based on one-dimensional Airy beam generated by single cylindrical lens," Opt. Commun. **393**, 11-16 (2017).
18. M. Avendaño-Alejo, L. Castañeda, and I. Moreno, "Properties of caustics produced by a positive lens: meridional rays," J. Opt. Soc. Am. A **27**, 2252-2260 (2010).

19. J. Yang, P. Twardowski, P. Gérard, Y. Duo, J. Fontaine, and S. Lecler, "Ultra-narrow photonic nanojets through a glass cuboid embedded in a dielectric cylinder," Opt. Express **26**, 3723-3731 (2018).
20. Y. Huang, Z. Zhen, Y. Shen, C. Min, and G. Veronis, "Optimization of photonic nanojets generated by multilayer microcylinders with a genetic algorithm," Opt. Express **27**, 1310-1325 (2019).
21. A. Taflove and S. Hagness, *Computational Electrodynamics: The Finite* Difference *Time Domain Method* (Artech House, 1998).
22. N. Yu, P. Genevet, M. Kats, F. Aieta, J. Tetienne, F. Capasso, and Z. Gaburro, "Light propagation with phase discontinuities: generalized laws of reflection and refraction," Science **334**, 333-337 (2011).
23. H. Yang, R. Trouillon, G. Huszka, and M. Gijs, "Super-resolution imaging of a dielectric microsphere is governed by the waist of its photonic nanojet," Nano Lett. **16**, 4862-4870 (2016).
24. G. Gu, L. Shao, J. Song, J. Qu, K. Zheng, X. Shen, Z. Peng, J. Hu, X. Chen, M. Chen, and Q. Wu, "Photonic hooks from Janus microcylinders," Opt. Express **27**, 37771-37780 (2019).
25. E. Palik, *Handbook of Optical Constants of Solids* (Academic Press, 1985).
26. Y. Zhang, C. Lo, J. Taylor, and S. Yang, "Replica molding of high-aspect-ratio polymeric nanopillar arrays with high fidelity," Langmuir **22**, 8595-8601 (2006).
27. C. Liu and K. Hsiao, "Direct imaging of optimal photonic nanojets from core-shell microcylinders," Opt. Lett. **40**, 5303-5306 (2015).
28. C. Liu and W. Lo, "Large-area super-resolution optical imaging by using core-shell microfibers," Opt. Commun. **399**, 104-111 (2017).
29. P. Ferrand, J. Wenger, A. Devilez, M. Pianta, B. Stout, N. Bonod, E. Popov, H. Rigneault, "Direct imaging of photonic nanojets," Opt. Express **16**, 6930-6940 (2008).